\documentclass[aps,prc,onecolumn,superscriptaddress]{revtex4}
\usepackage[latin1]{inputenc}
\usepackage{bm}
\usepackage{epsfig}
\usepackage{amsthm}
\usepackage{amsfonts}
\usepackage{float}
\usepackage{amsmath,amssymb}
\usepackage{color}
\usepackage{slashed}
\usepackage{relsize}
\usepackage[toc,page]{appendix} 
\usepackage{wrapfig}
\usepackage{graphicx}
\usepackage{dcolumn}
\usepackage{bm}
\newcommand{\be}{\begin{equation}} 
\newcommand{\ee}{\end{equation}} 
\newcommand{\bea}{\begin{eqnarray}} 
\newcommand{\eea}{\end{eqnarray}} 
\newcommand{\nn}{\nonumber} 
\newcommand{\mintedim}[2]{{\int\kern-0.50em\mbox{{\small$\mathop{\frac{\mbox{{\small${\rm d^{#2}}\vect{#1}$}}}{\mbox{{\small$(2\pi)^{#2}$}}}}$}}\ }} 
\newcommand{\inteonedim}[1]{{\int_0^\infty\kern-1em\mbox{{\small${\rm d}{#1}$}}}} 
\newcommand{\vect}[1]{\bm{#1}} 

\begin{document}
\title{A harmonic oscillator in nonadditive statistics and a novel transverse momentum spectrum in high-energy collisions}

\author{Trambak Bhattacharyya}
\email{trambak.bhattacharyya@ujk.edu.pl}
\affiliation{Institute of Physics, Jan Kochanowski University, Kielce 25-406, Poland}

\author{Maciej Rybczy\'{n}ski}
\email{maciej.rybczynski@ujk.edu.pl}
\affiliation{Institute of Physics, Jan Kochanowski University, Kielce 25-406, Poland}

\author{Grzegorz Wilk}
\email{Grzegorz.Wilk@ncbj.gov.pl}
\affiliation{National Centre For Nuclear Research, Pasteura 7, Warsaw 02-093, Poland}

\author{Zbigniew W{\l}odarczyk}
\email{zbigniew.wlodarczyk@ujk.edu.pl}
\affiliation{Institute of Physics, Jan Kochanowski University, Kielce 25-406, Poland}
\begin{abstract} 
It is widely observed that particles produced in high-energy collisions follow a power-law distribution. One such power-law distribution used extensively in the phenomenological
studies owes its origin to nonadditive statistics proposed by C. Tsallis. In this article, we derive a novel nonadditive generalization of the conventional Bose-Einstein distribution
using a single-mode harmonic oscillator. The approach taken in this paper eliminates the need of a regularization procedure proposed in previous works.
We observe that the spectra of the bosonic particles like the pions and kaons  produced in high-energy collisions 
are well-described by the nonadditive bosonic distribution derived in this paper.
\end{abstract}

\maketitle
\section{Introduction}
Transverse momentum spectra of the hadrons produced in high-energy collisions are important experimental observables that provide information on dynamics of the system and the freeze-out parameters like temperature. Hence, it is important to understand them from a theoretical perspective and these efforts constitute an active field of research. Because of a large number of particles produced in these collisions, methods of statistical mechanics in describing particle spectra can be applied. However, the conventional Boltzmann-Gibbs statistics, that yields exponential distributions, is unable to describe experimental data at larger momentum. For example, for the $\pi^+$ spectra in a p-p collision at $\sqrt{s}=900$ GeV measured by the ALICE collaboration \cite{alicepi},  the experimental spectra obtained in high-energy collisions start deviating from being exponential at a transverse momentum around 0.5 GeV. On the other hand, power-law distributions successfully describe the transverse momentum distributions of such particles \cite{TsCMS,TsALICE,abcp,maciejepjc} up to a large momentum. These references utilize a certain class of power-law distributions that owes its origin to nonadditive statistics proposed by C. Tsallis \cite{Tsal88}. 

Nonadditive statistics is a generalization of the Boltzmann-Gibbs statistics for systems having fluctuations and long-range correlation. Such a generalization often features the following quasi-exponential function:
\bea
\exp_q \left(-\frac{\varepsilon_p}{T} \right) = \left(1+\frac{q-1}{T}\varepsilon_p\right)^{-\frac{1}{q-1}},
\label{qexp}
\eea
where $q$ is the entropic parameter, $\varepsilon_p=\sqrt{p^2+m^2}$ is the single particle energy of a particle of the mass $m$ and 3-momentum $\vec{p}$ ($|\vec{p}|\equiv p$), and $T (\equiv \beta^{-1})$ is temperature. When $q$ approaches 1, Eq.~\eqref{qexp} approaches toward the conventional exponential function. 

The $q$ parameter is shown to be connected to the relative variance in thermodynamic quantities (e.g. temperature) \cite{Wilk00,Wilk09}. The $q$ parameter can also be computed in terms of the parameters in quantum chromodynamics like $N_c$ (no. of colours) and $N_f$ (no. of flavours) \cite{deppmanprdq} by studying the scaling properties of the Yang-Mills theory. The entropic parameter has been utilized to estimate the relaxation and correlation times of a hadronizing system \cite{maciejprd} or to propose a generalized Boltzmann transport equation whose stationary state solution is given by the $q$-exponential function \cite{lavagnopla,wilkosada,Biro:2012ix}. There are many instances where natural systems exhibit behaviour that is better described by generalized statistical mechanics, of which nonadditive statistics is a prominent example.

\begin{figure}[!htb]
\vspace{0.4in}
\minipage{0.45\textwidth}
\begin{center}
\hspace{-0.4in}
\includegraphics[width=\linewidth]{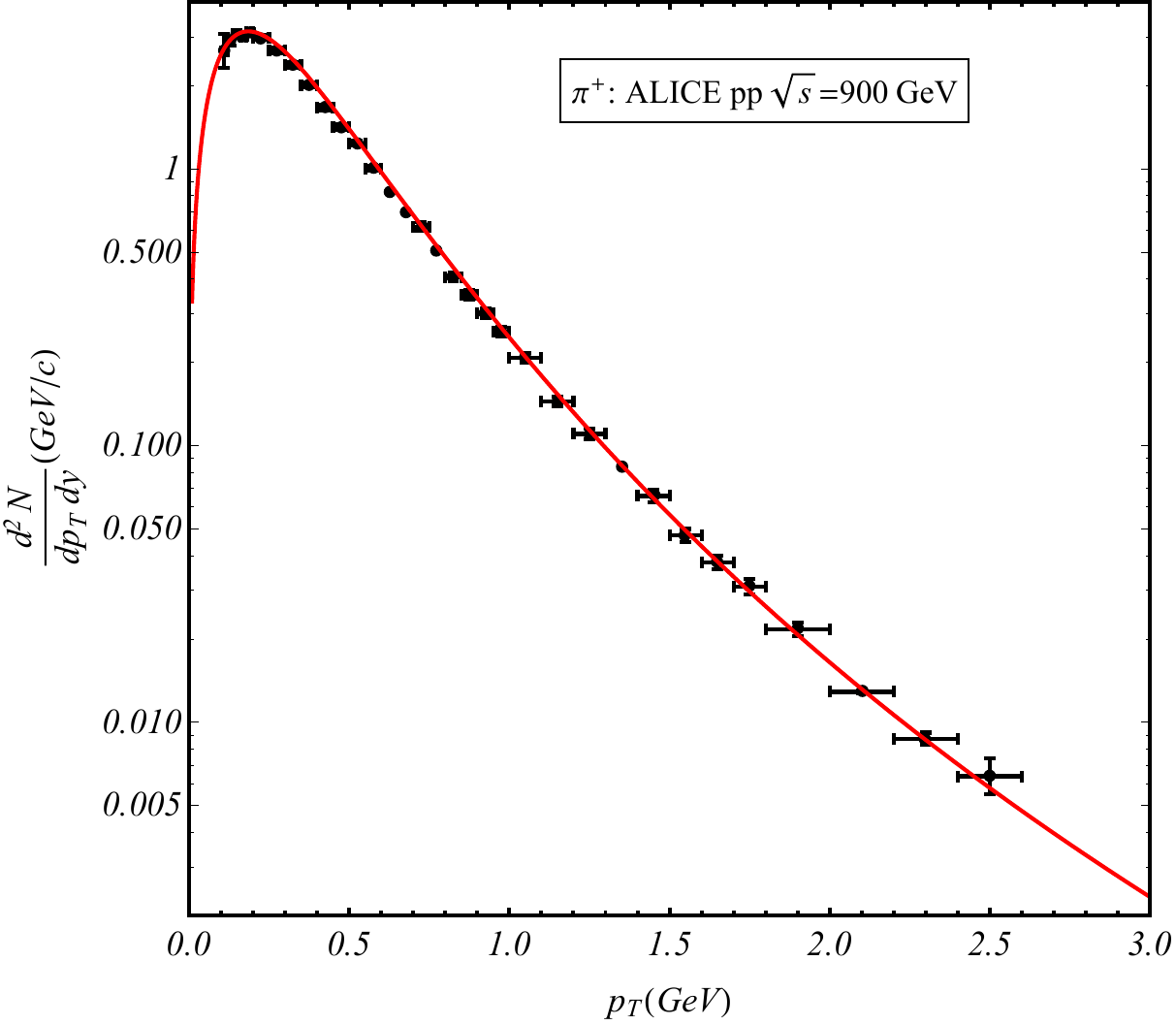}
\caption{Description of $\pi^+$ spectrum provided in Ref.~\cite{alicepi} using Eq.~\eqref{tsallisBEspd}. The parameter values are: $q=0.8795 \pm 0.004391$, $T=0.0883 \pm 0.002957$ GeV, $R = 3.873 \pm 0.1063$ fm, $\chi^2$/NDF = 2.667/30.}\label{pi900}
\end{center}
\endminipage\hfill
\minipage{0.45\textwidth}
\vspace*{-0.2cm}
\hspace*{-2cm}
\includegraphics[width=\linewidth]{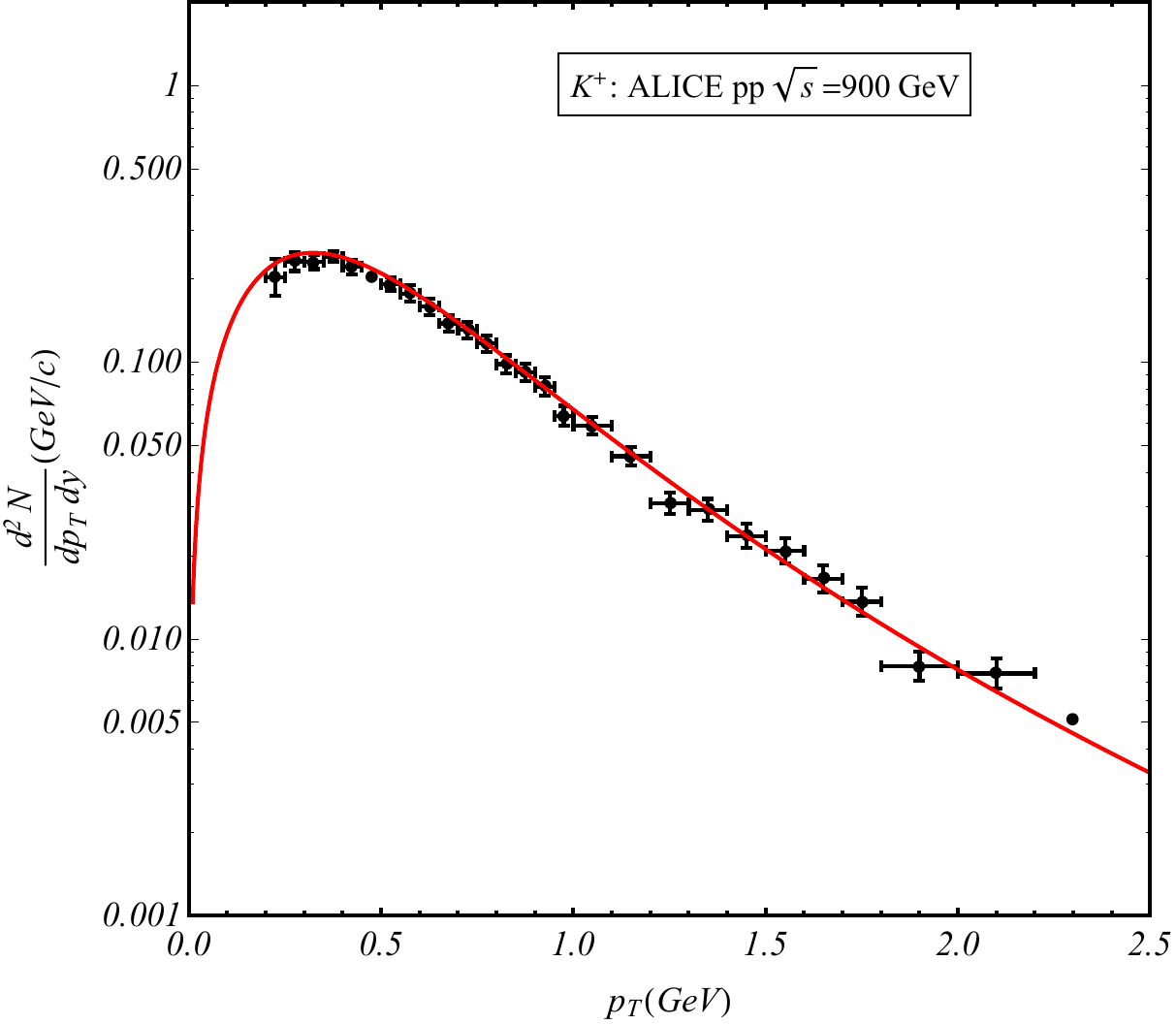}
\caption{Description of $K^+$ spectrum provided in Ref.~\cite{alicepi} using Eq.~\eqref{tsallisBEspd}: The parameter values are: $q=0.853 \pm 0.01213$, $T=0.0657 \pm 0.01433$ GeV, $R = 3.994 \pm 1.083$ fm, $\chi^2$/NDF = 5.357/24.}\label{k900}
\endminipage\hfill
\end{figure}

It is important to establish a connection between the single-particle distribution used in the literature and the fundamental approaches of physics like statistical mechanics considered in the present work. It has been shown that the phenomenological nonadditive distributions can be derived \cite{TBParvan1,TBParvan2} by extremizing nonadditive entropy proposed by Tsallis \cite{Tsal88}. However, while deriving classical and quantum distributions, divergence was encountered and a regularization scheme had to be proposed. In this article, we take a different approach in which we derive a novel bosonic nonadditive single-particle distribution considering a single-mode harmonic oscillator. This approach eliminates the need of a regularization scheme. We use this nonadditive bosonic distribution to model particle transverse momentum distribution and compare our results with experimentally observed spectra. Highlights of the present work are as follows:
\begin{itemize}
\item As far as our knowledge goes, the nonadditive transverse momentum spectrum in Eq.~\eqref{t1spectra1} has been derived for the first time in the literature considering a single-mode harmonic oscillator. 
\item While the conventional Bose-Einstein distribution describes only very low momentum data, the nonadditive transverse momentum spectrum derived in Eq.~\eqref{t1spectra1} describes experimental data well for the whole range of transverse momenta considered in our analyses (Figs. \ref{pi900} and \ref{k900}). 
\item Our approach eliminates the need of regularization of transverse momentum spectra.
\end{itemize}
The rest of the article will be devoted to describing the mathematical model, exploring different limits of our result and comparing it with experimental data.

\section{Mathematical model from nonadditive statistical mechanics}

\subsection{Equilibrium set of probabilities}
Nonadditive statistical mechanics is based on the following definition of entropy~\cite{Tsal88},
\be
S =\mathlarger{ \mathlarger{ \sum}}\limits_{i} \frac{p_{i}^{q}-p_{i}}{1-q},
\label{t1entropy}
\ee
where $q$ is a real parameter, and the probabilities of micro-states $\{p_i\}$ follow the normalization condition,
\be
\varphi=\sum\limits_{i} p_i -1=0.
\label{probnorm}
\ee
The definition of average expectation values in the normalized (or the first) scheme
is given by \cite{Tsal98},
\be \label{t1av}
\langle Q \rangle = \sum\limits_{i} p_i Q_i .
\ee

The thermodynamic potential $\Omega$ of the grand canonical ensemble can be written as,
\begin{eqnarray}
 \Omega &=& \langle H \rangle -TS-\mu \langle N \rangle \nonumber \\
  &=&  \sum\limits_{i}  p_{i} \left[E_{i}-\mu N_{i} - T \frac{p_{i}^{q-1}-1}{1-q}\right],
  \label{t1pot}
\end{eqnarray}
where $\langle H \rangle=\sum_{i}  p_{i} E_{i}$ is the mean energy of the system, $\langle N \rangle=\sum_{i}  p_{i} N_{i}$ is the mean number of particles, and $E_{i}$ and $N_{i}$ are the energy and number of particles, respectively, in the $i$-th microscopic state of the system. The set of equilibrium probabilities $\{p_{i}\}$ can be found from local extremization of the thermodynamic potential $\Omega$ (subjected to probability normalization constraint) by the method of the Lagrange multipliers (see, for example, Ref.~\cite{Jaynes2}). In terms of a modified
potential $\Phi$, defined below, the equilibrium set of probabilities can be found from the second of the following equations:
\begin{eqnarray}\label{phi}
 \Phi &=& \Omega - \lambda \varphi,  \\ \label{delphidelpi}
  \frac{\partial \Phi}{\partial p_{i}} &=& 0,
\end{eqnarray}
where $\lambda$ is an arbitrary real constant. 

Substituting Eqs.~\eqref{probnorm} and \eqref{t1av} into Eqs.~\eqref{phi} and \eqref{delphidelpi}, we obtain the  equilibrium probabilities of the grand canonical ensemble (for the normalized statistics) as~\cite{TBParvan2}
\begin{equation}
p_{i} = \left[1+\frac{q-1}{q}\frac{\Lambda-E_{i}+\mu N_{i}}{T}\right]^{\frac{1}{q-1}},
\label{t1prob}
\end{equation}
subjected to probability normalization,
\begin{equation}
    \mathlarger{\sum}\limits_{i} \left[1+\frac{q-1}{q}\frac{\Lambda-E_{i}+\mu N_{i}}{T}\right]^{\frac{1}{q-1}}=1,
\label{probnormint}
\end{equation}
where $\Lambda\equiv \lambda-T$ and $\partial E_{i}/\partial p_{i}=\partial N_{i}/\partial p_{i}=0$. $\Lambda$ is related to the partition function that also helps define an effective
temperature. In the Gibbs limit $q\to 1$, the probability $p_{i}=\exp[(\Lambda_{\text{G}}-E_{i}+\mu N_{i})/T]$, where $\Lambda_{\text{G}}=-T\ln Z_{\text{G}}$ is the thermodynamic potential of the grand canonical ensemble and $Z_{\text{G}}=\sum_{i} \exp[-(E_{i}-\mu N_{i})/T]$ is the partition function. 

\subsection{Nonadditive average represented in terms of Boltzmann-Gibbs average}
Using the integral representation of the gamma functions~\cite{Abramowitz} for $q<1$, probability normalization and average values can be rewritten as~\cite{TBParvan2},  

\begin{subequations}
\bea
\label{t1probintrep}
p_{i} &=& \frac{1}{\Gamma\left(\frac{1}{1-q}\right)} \int\limits_{0}^{\infty} t^{\frac{q}{1-q}} e^{-t\left[1+\frac{q-1}{q}\frac{\Lambda-E_{i}+\mu N_{i}}{T}\right]} dt;
\nn\\
\sum_{i} p_i &=& \frac{1}{\Gamma\left(\frac{1}{1-q}\right)} \int\limits_{0}^{\infty} t^{\frac{q}{1-q}} e^{-t\left[1+\frac{q-1}{q}\frac{\Lambda-\Omega_{\text{G}}\left(\beta'\right)}{T}\right]} dt  = 1.
\eea
\bea
\langle Q \rangle &=& 
\frac{1}{\Gamma\left(\frac{1}{1-q}\right)} \int\limits_{0}^{\infty} t^{\frac{q}{1-q}} 
e^{ -t \left[1+\frac{q-1}{q}\frac{\Lambda}{T} \right]}
Z_{\text{G}}\left(\beta'\right)
\langle Q \rangle_{\text{G}} (\beta') dt,
\label{t1avintrep}
\eea
\end{subequations}
where
\begin{eqnarray}
\label{1a}
 \Omega_{\text{G}}\left(\beta'\right) &=& -\frac{1}{\beta'} \ln Z_{\text{G}}\left(\beta'\right);~~
Z_{\text{G}}\left(\beta'\right) = \sum\limits_{i} e^{-\beta'(E_{i}-\mu N_{i})}; \nn\\
\langle Q \rangle_{\text{G}}\left(\beta'\right) &=& \frac{1}{Z_{\text{G}}\left(\beta'\right)} \sum\limits_{i} Q_{i} e^{-\beta'(E_{i}-\mu N_{i})};
\nn\\
\text{and} ~~~~
\beta' &=& t(1-q)/qT.
\end{eqnarray}

The main results of this section are Eqs.~\eqref{t1probintrep} and \eqref{t1avintrep} that will be used to calculate nonadditive single particle distributions ($\langle n_{p\sigma} \rangle$) in terms of Boltzmann-Gibbs single particle distributions ($\langle n_{p\sigma} \rangle_{\text{G}}$). $n_{p\sigma}$ is the number of particles (of the mass $m$ and energy $\varepsilon_p$) in a micro-state $i$ with three-momentum $p=\sqrt{\varepsilon_p^2-m^2}$, such that $E_i=\sum n_{p\sigma} \varepsilon_p$ and $N_i=\sum n_{p\sigma}$ ($\sigma$ represents any quantum number like spin, for example).

\subsection{Calculating nonadditive single-particle distributions}
Using Eq.~\eqref{t1avintrep}, we can express the nonadditive single-particle distributions in terms of the Boltzmann-Gibbs single-particle distributions through the following
integral,

\bea
\langle n_{p\sigma} \rangle 
&=& 
\frac{1}{\Gamma\left(\frac{1}{1-q}\right)} 
\int\limits_{0}^{\infty} 
t^{\frac{q}{1-q}} 
e^{-t\left[1+\frac{q-1}{q}\frac{\Lambda}{T}\right]} 
~
Z_{\text{G}}\left(\beta'\right) 
~\langle n_{p\sigma} \rangle_{\text{G}} (\beta') ~dt.
\label{t1spdintrep}
\eea
Eq.~\eqref{t1spdintrep}, that relates mean values in the nonadditive statistics with those in the Boltzmann-Gibbs statistics, is referred to as the Hilhorst integral transformation \cite{curilef} that has been frequently used in the literature \cite{TBParvan1,TBParvan2,prato}.
In this integral, $\Lambda$ is calculated from probability normalization given by Eq.~\eqref{t1probintrep} for a given set of $q$ and $T$. We are able to find
a closed analytical form of Eq.~\eqref{t1spdintrep} for a single-mode harmonic oscillator as shown below.

In the Boltzmann-Gibbs statistics, the (normal ordered) partition function for a single-mode harmonic oscillator (with frequency $\varepsilon_p$) is given by (chemical potential is set equal to zero),
\bea
Z_{\text{G}} (\beta) =  \frac{1}{1-e^{-\beta \varepsilon_p}},
\label{zshobg}
\eea 
and it yields the Boltzmann-Gibbs Bose-Einstein single-particle distribution,
\bea
\langle n_{p\sigma} \rangle_{\text{G}}(\beta) &=& -\frac{1}{\beta} \frac{\partial \ln Z_{\text{G}}}{\partial \varepsilon_p} \nn\\
&=& \frac{1}{e^{\beta \varepsilon_p}-1}.
\label{spdshobg}
\eea

In what follows, we generalize the Boltzmann-Gibbs Bose-Einstein distribution obtained in Eq.~\eqref{spdshobg} for nonadditive statistics by putting Eqs.~\eqref{zshobg} and \eqref{spdshobg} in Eq.~\eqref{t1spdintrep}:
\bea
&&\langle n_{p\sigma} \rangle = \frac{1}{\Gamma\left(\frac{1}{1-q}\right)} 
\int\limits_{0}^{\infty} 
t^{\frac{q}{1-q}} 
e^{-t\left[1+\frac{q-1}{q}\frac{\Lambda}{T}\right]} 
~
e^{-\beta'\Omega_{\text{G}}(\beta')}
~\langle n_{p\sigma} \rangle_{\text{G}} (\beta') ~dt
\nn\\
&=& 
\frac{1}{\Gamma\left(\frac{1}{1-q}\right)} 
\int\limits_{0}^{\infty} 
t^{\frac{q}{1-q}} 
e^{-t\left[1+\frac{q-1}{q}\frac{\Lambda}{T}\right]} 
~
\frac{1}{1-e^{-\beta' \varepsilon_p}}
~\frac{1}{e^{\beta' \varepsilon_p}-1} ~dt
\nn\\
&=& 
\frac{1}{\Gamma\left(\frac{1}{1-q}\right)} 
\int\limits_{0}^{\infty} 
t^{\frac{q}{1-q}} 
e^{-t\left[1+\frac{q-1}{q}\frac{\Lambda}{T}\right]} 
~
\sum_{n=0}^{\infty} e^{-n \beta'\varepsilon_p }
~\left[\sum_{r=0}^{\infty}e^{- r \beta'\varepsilon_p} -1\right]~dt
\nn\\
&=& 
\frac{1}{\Gamma\left(\frac{1}{1-q}\right)} 
\sum_{n=0}^{\infty} \sum_{r=0}^{\infty}
\int\limits_{0}^{\infty} 
t^{\frac{q}{1-q}} 
e^{-t\left[1+\frac{q-1}{q}\frac{\Lambda}{T}\right]} 
~
 e^{-(n+r) \beta'\varepsilon_p }~dt
 -
 \frac{1}{\Gamma\left(\frac{1}{1-q}\right)} 
\sum_{n=0}^{\infty}
\int\limits_{0}^{\infty} 
t^{\frac{q}{1-q}} 
e^{-t\left[1+\frac{q-1}{q}\frac{\Lambda}{T}\right]} 
~
 e^{-n \beta'\varepsilon_p }~dt
 \nn\\
 &=& 
\mathlarger{\sum}_{n=0}^{\infty} \mathlarger{\sum}_{r=0}^{\infty}
\frac{1}{\left ( 1+\frac{1-q}{q T} \left[ (n+r)\varepsilon_p-\Lambda \right] \right)^{\frac{1}{1-q}}}
-
\mathlarger{\sum}_{n=0}^{\infty}
\frac{1}{\left ( 1+\frac{1-q}{q T}[n \varepsilon_p-\Lambda] \right)^{\frac{1}{1-q}}}
 \nn\\
 &=& 
\left( \frac{qT}{(1-q)\varepsilon_p} \right)^{\frac{1}{1-q}}
\left[
\zeta \left( \frac{q}{1-q}, \frac{qT}{(1-q)\varepsilon_p} - \frac{\Lambda}{\varepsilon_p}  \right)
+
\left( 1- \frac{qT}{(1-q)\varepsilon_p} + \frac{\Lambda}{\varepsilon_p} \right) \zeta \left( \frac{1}{1-q}, \frac{qT}{(1-q)\varepsilon_p} - \frac{\Lambda}{\varepsilon_p}  \right)
\right]
\nn\\
&&-
\left( \frac{qT}{(1-q)\varepsilon_p} \right)^{\frac{1}{1-q}} 
\zeta \left( \frac{1}{1-q}, \frac{qT}{(1-q)\varepsilon_p} - \frac{\Lambda}{\varepsilon_p}  \right)
\nn\\
&=& \left( \frac{qT}{(1-q)\varepsilon_p} \right)^{\frac{1}{1-q}} 
\left[
\zeta \left( \frac{q}{1-q}, \frac{qT}{(1-q)\varepsilon_p} - \frac{\Lambda}{\varepsilon_p}  \right)
-
\left( \frac{qT}{(1-q)\varepsilon_p} - \frac{\Lambda}{\varepsilon_p} \right) \zeta \left( \frac{1}{1-q}, \frac{qT}{(1-q)\varepsilon_p} - \frac{\Lambda}{\varepsilon_p}  \right)
\right]
\label{tsallisBEspd}
\eea

Using a similar procedure, the probability normalization, Eq.~\eqref{t1probintrep}, can be written as:
\bea
\left( \frac{qT}{(1-q)\varepsilon_p} \right)^{\frac{1}{1-q}}  \zeta \left( \frac{1}{1-q}, \frac{qT}{(1-q)\varepsilon_p} - \frac{\Lambda}{\varepsilon_p}  \right) = 1.
\eea
Solving the above equation gives us the value of $\Lambda$. In the above equations $\zeta(s,a)$ is the Hurwitz zeta function defined as follows:
\bea
\zeta(s,a) = \sum_{n=0}^{\infty} \frac{1}{(a+n)^s}, \forall n \ni a+n \neq 0, \Re(s)>1.
\eea
We also use the following identity,
\bea
\sum_{m,n=0}^{\infty} \frac{1}{(b+m+n)^s} = \zeta(s-1,b) + (1-b) \zeta(s,b), \forall b>0.
\eea

\subsection{Single particle distribution: some observations}
\begin{figure}[!htb]
\minipage{0.45\textwidth}
\begin{center}
\hspace{-0.4in}
\includegraphics[width=\linewidth]{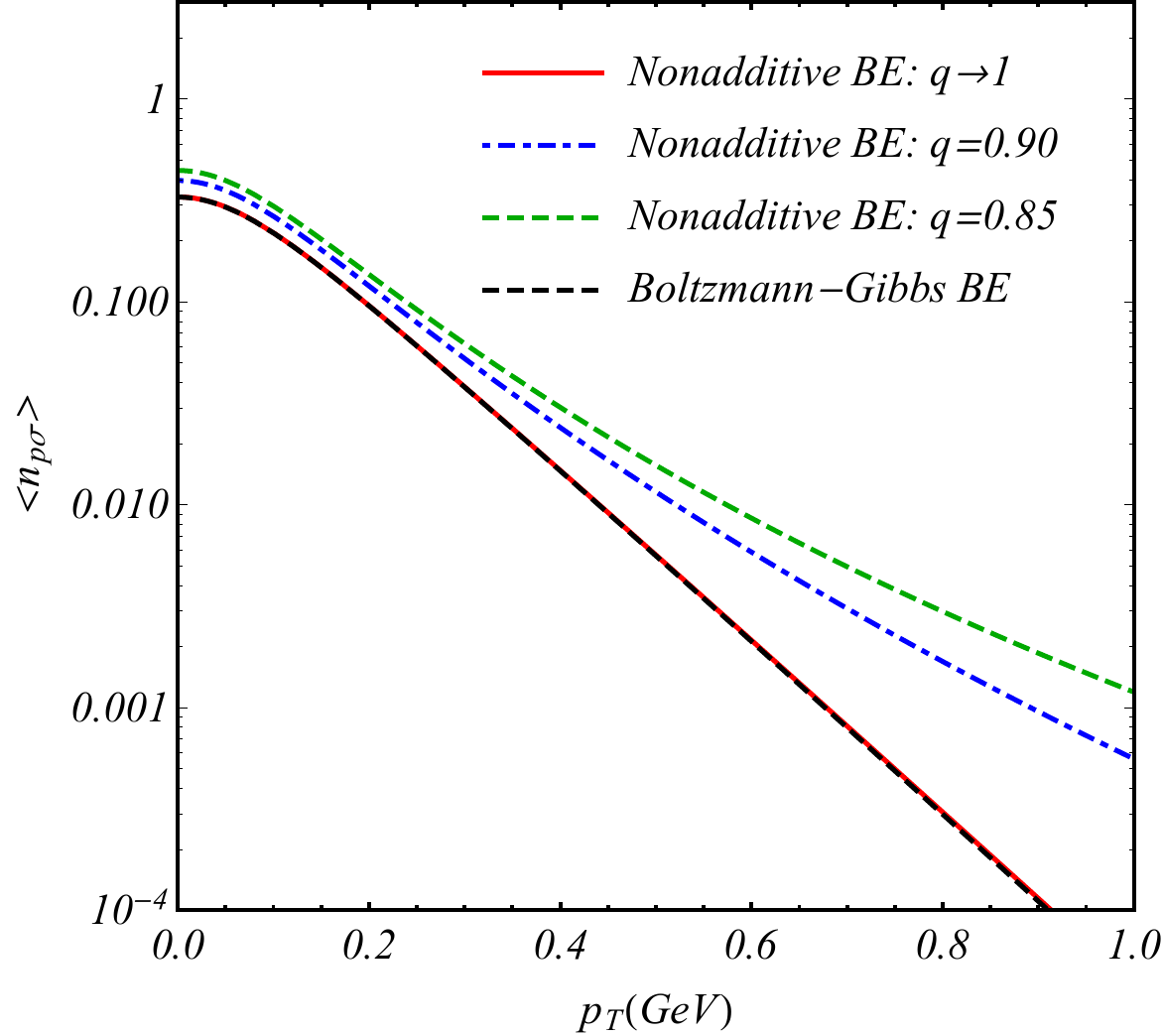}
\caption{Variation of nonadditive BE distribution (Eq.~\eqref{tsallisBEspd}) for a particle having the mass of the pions (0.14 GeV) and temperature $T=0.1$ GeV for different
entropic parameter $q$.}
\label{TsBEtoBGBE}
\end{center}
\endminipage\hfill
\minipage{0.44\textwidth}
\vspace*{-0.4cm}
\hspace*{-2cm}
\includegraphics[width=\linewidth]{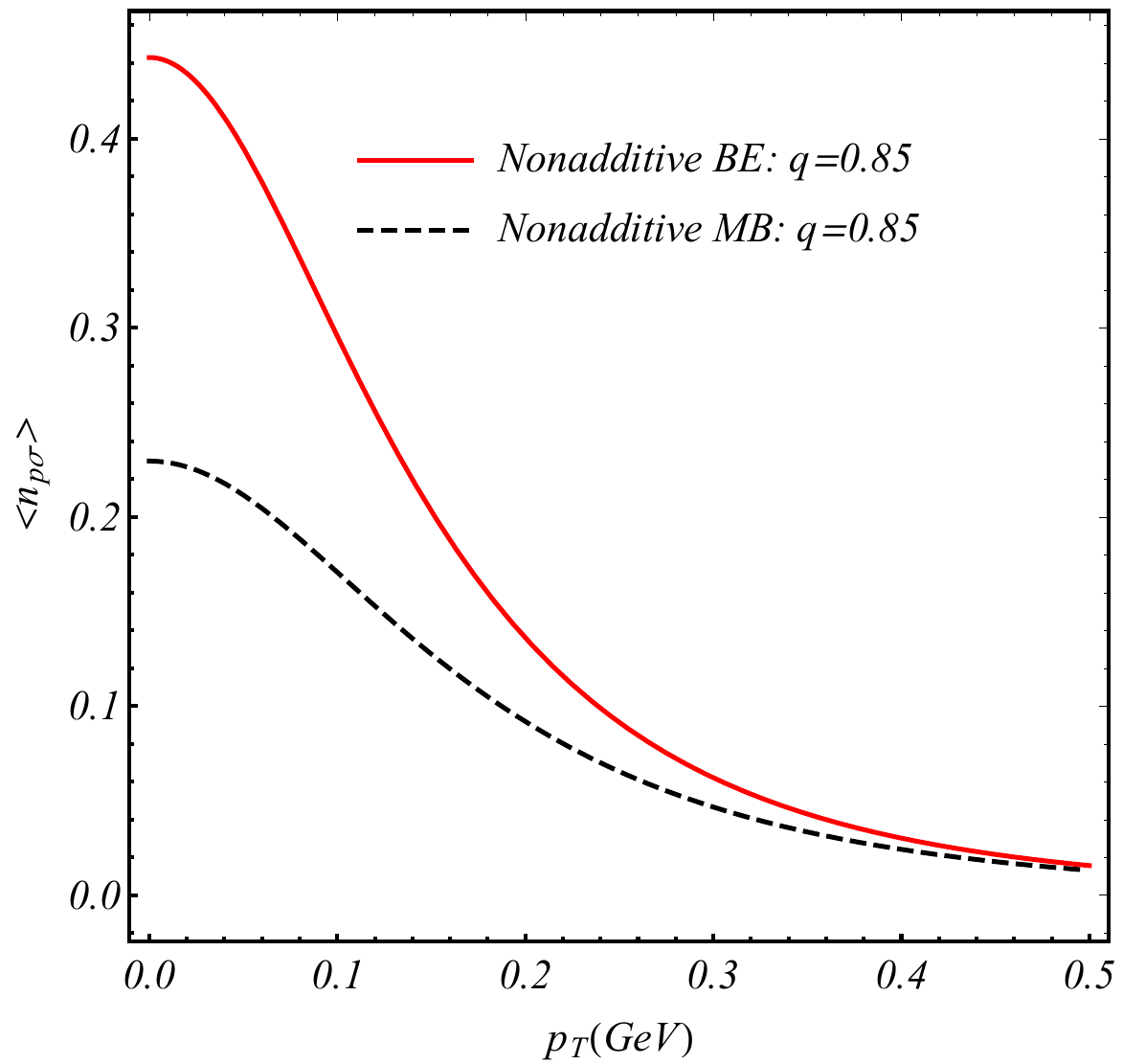}
\hspace{-1cm}
\caption{Nonadditive BE distribution in Eq.~\eqref{tsallisBEspd} ($m=0.14$ GeV, $T$ = 0.1 GeV) approaches classical nonadditive distribution when energy well surpasses temperature.}
\label{TsBEtoMB}
\endminipage\hfill
\end{figure}

When $q$ approaches 1, Eq.~\eqref{tsallisBEspd} approaches the conventional Boltzmann-Gibbs Bose-Einstein (BGBE) distribution, as seen in Fig.~\ref{TsBEtoBGBE}. The emergence of the BGBE distribution can also be understood by taking
the $q\rightarrow1$ limit of the fifth line of Eq.~\eqref{tsallisBEspd}:
\bea
\mathlarger{\sum}_{n=0}^{\infty} \mathlarger{\sum}_{r=0}^{\infty}
\frac{1}{\left ( 1+\frac{1-q}{q T} \left[ (n+r)\varepsilon_p-\Lambda \right] \right)^{\frac{1}{1-q}}}
-
\mathlarger{\sum}_{n=0}^{\infty}
\frac{1}{\left ( 1+\frac{1-q}{q T}[n \varepsilon_p-\Lambda] \right)^{\frac{1}{1-q}}} 
\xrightarrow{q \rightarrow 1}
\frac{\exp[\beta(\varepsilon_p+\Lambda_{\text{G}})]}{(\exp(\beta \varepsilon_p)-1)^2}
=
 \frac{1}{\exp(\beta \varepsilon_p)-1},
 \nn\\
\eea
where we use Eq.~\eqref{zshobg} and the relationships below Eq.~\eqref{probnormint} to find that $\exp(\beta \Lambda_{\text{G}})=Z_{\text{G}}^{-1}=1-\exp(-\beta \epsilon_p)$. 

We also observe in Fig.~\ref{TsBEtoMB} that as the single-particle energy begins surpassing other energy scales, the nonadditive BE distribution in Eq.~\eqref{tsallisBEspd} starts approaching the classical nonadditive Maxwell-Boltzmann (MB) distribution given below:
\bea
\langle n_{p\sigma} \rangle \xrightarrow{\varepsilon_p \gg T, \Lambda} \left( 1+ \frac{1-q}{qT} \varepsilon_p \right)^{\frac{1}{q-1}}.
\label{tsspdmb}
\eea
The distribution above is dual to the widely-used phenomenological nonadditive distribution:
\bea
\left( 1+ \frac{1-q}{qT} \varepsilon_p \right)^{\frac{1}{q-1}} \xrightarrow{q \rightarrow 1/q'} \left( 1+ \frac{q'-1}{T} \varepsilon_p \right)^{-\frac{q'}{q'-1}}.
\eea
The classical distribution can also be obtained by imposing the factorization approximation \cite{hasegawa} on Eq.~\eqref{tsallisBEspd} in high energy limit. The factorization approximation 
(as well as assuming $\varepsilon_p \gg T, \Lambda$) amounts to making the following substitution (for a summation index $N \in \mathbb{Z}^{\geq}$):
\bea
\left ( 1+\frac{1-q}{q T}[N \varepsilon_p-\Lambda] \right)^{-\frac{1}{1-q}} \approx \left ( 1+\frac{1-q}{q T}\varepsilon_p \right)^{-\frac{N}{1-q}}.
\label{factapp}
\eea
Using Eq.~\eqref{factapp} and performing the series summation in Eq.~\eqref{tsallisBEspd} lead us to:
\bea
\langle n_{p\sigma} \rangle _{\text{F}} \approx \left(1+\frac{1-q}{qT} \varepsilon_p \right)^{\frac{1}{q-1}},
\eea
where `F' stands for a factorized single-particle distribution. In the lower energy region, $\langle n_{p\sigma} \rangle _{\text{F}}$ approaches being dual ($q\leftrightarrow1/q'$) to the phenomenological nonadditive Bose-Einstein distribution used in the literature \cite{tsq}.

\section{Results and discussion}
There have been many attempts to utilize the single-particle distributions to describe particle production in high-energy collisions. 
Experimentally observed transverse momentum distributions can be calculated from single-particle distributions $\langle n_{p\sigma} \rangle$ in the following way:

\bea
\frac{d^2N}{dp_{\text{T}}dy} = \frac{gV}{(2\pi)^3} \int\limits_{0}^{2\pi} d\phi ~p_{\mathrm{T}} \varepsilon_p \langle n_{p\sigma}\rangle,
\label{t1spectra}
\eea
where $\phi$ is the azimuthal angle, $g$ is the degeneracy factor, $p_{\text{T}}$ is transverse momentum, and $V=4\pi R^3/3$ ($R$: radius) is the volume. Some of the approaches utilize the exponential Boltzmann-Gibbs (for low momenta), or the phenomenological nonadditive power-law distributions. However, in this work, we utilize the single-particle distribution derived in Eq.~\eqref{tsallisBEspd} and obtain the following expression:
\bea
\frac{d^2N}{dp_{\text{T}}dy} &=& \frac{gV}{(2\pi)^2} p_{\mathrm{T}}  m_{\mathrm{T}} \cosh(y) \nn\\
&&\times \left( \frac{qT}{(1-q)\varepsilon_p} \right)^{\frac{1}{1-q}} 
\left[
\zeta \left( \frac{q}{1-q}, \frac{qT}{(1-q)\varepsilon_p} - \frac{\Lambda}{\varepsilon_p}  \right)
-
\left( \frac{qT}{(1-q)\varepsilon_p} - \frac{\Lambda}{\varepsilon_p} \right) \zeta \left( \frac{1}{1-q}, \frac{qT}{(1-q)\varepsilon_p} - \frac{\Lambda}{\varepsilon_p}  \right)
\right],
\label{t1spectra1}
\nn\\
\eea
where single-particle energy (of a particle of the mass $m$ and momentum $\vec{p} \equiv \{\vec{p_{\text{T}}},p_z\}$) $\varepsilon_p=\sqrt{\vec{p}^2+m^2}$ is parameterized in terms
of transverse mass $m_{\text{T}}=\sqrt{p_{\text{T}}^2+m^2}$ and rapidity $y$ such that $\varepsilon_p = m_{\text{T}} \cosh(y)$. Eq.~\eqref{t1spectra1} is the main result of our paper that
we use to study particle spectra produced in high-energy collisions.

%

 However, it is noteworthy that Eq~\eqref{t1spectra1} is an approximation of the generalized Bose-Einstein distribution within the scope of nonadditive statistics. In the most generalized scenario, one has to evaluate a summation over all the micro-states $i$, whereas choosing a single-mode harmonic oscillator leads to a simplification by removing all other single-particle states except $p\sigma$. This simplification also is essentially based on the factorizability assumption imposed on Eqs.~\eqref{t1prob} and \eqref{probnormint}. Factorization approximation in the context of nonadditive statistics has been studied for a long time (see for example \cite{buyufact1,buyufact2}). Validity of such an approximation has been the subject matter of several works and requires careful analysis like that in Ref.~\cite{tirnakli} that concludes that, for simple bosonic systems and for $q$ values within the known limits, factorization approximation may be useful. 
Computing the summation over all the micro-states is a challenging task in the context of nonadditive statistics and there are examples in the literature (e.g. \cite{tsaapre,tirnakli,wongchaos,lenzipla}) that discuss approximated results valid for some types of systems. Our present study, that utilizes the standard linear mean values unlike some previous works, is aligned with this line of investigation.

\section{Summary, outlook, and conclusion}
To summarize, we have derived a novel nonadditive bosonic single-particle distribution by extremizing nonadditive entropy and considering a single-mode harmonic oscillator that simplifies the summation over all micro-states. By utilizing the methods of nonadditive statistical mechanics to study a single-mode simple harmonic oscillator, we avoid diverging results reported in previous works. The nonadditive distribution well describes spectra of the bosonic particles like the pions and kaons produced in high-energy collisions, while the conventional Bose-Einstein spectrum deviates from experimental data at around $p_{\text{T}}\sim$ 0.5 GeV. 
The bosonic distribution in Eq.~\eqref{tsallisBEspd} approaches the conventional Bose-Einstein distribution in the limit $q\rightarrow1$. The factorization approximation of the distribution in high-energy limit ($\epsilon_p \gg T,~\Lambda$) results in the classical nonadditive distribution. 

A single-particle distributions similar to (but not exactly the same as) Eq.~\eqref{tsallisBEspd} also appears in the thermal Green's function in nonadditive statistics \cite{AbeEPJB}. However, Ref.~\cite{AbeEPJB} considers a different definition of mean values given by escort distribution and a proper comparison can be made only when such a definition is considered in the present work as well. Using the escort probabilities, it is possible to evaluate thermodynamic quantities of a system of harmonic oscillators of different frequencies \cite{IshiharaEPJB,Ishiharaarxiv}. It will be interesting to study if the result of the present work can be reproduced within the framework of these references. It is also worth mentioning that the classical limit of Eq.~\eqref{tsallisBEspd} appears as the stationary solution of the nonadditive Boltzmann transport equation that considers a generalization of the `molecular chaos hypothesis' \cite{wilkosada}. 

Extension to fermions may be performed by considering a fermionic oscillator. The Hilbert space of a fermionic harmonic oscillator contains only two states ($|0\rangle$ and $|1\rangle$). By using the BG partition function of a single-mode fermionic oscillator, given by $Z_{\text{G}}(\beta') = 1+ \exp(-\beta'\epsilon_p)$, in Eq. \eqref{t1spdintrep}, one may be able to derive that the fermionic distribution is given by $\left(1+\frac{1-q}{qT} (\epsilon_p-\Lambda)\right)^{-\frac{1}{1-q}}$ which in the limit $q\rightarrow 1$ (BG limit) yields $\exp(-\beta(\epsilon_p-\Lambda_{\text{G}})$. Since $\exp(\beta \Lambda_{\text{G}})=Z_{\text{G}}(\beta)^{-1}$, the nonadditive distribution for fermions approach the Boltzmann-Gibbs Fermi-Dirac distribution in the limit $q\rightarrow 1$.
Given the bosonic distribution in Eq.~\eqref{tsallisBEspd}, it may be possible to estimate thermal mass and express strong coupling $\alpha_{\text{s}}$ in terms of temperature and the entropic parameter $q$ \cite{SukanyaEPJC}. Such a parameterization of the strong coupling may be utilized to improve the description of $\alpha_{\text{s}}$ at low energies \cite{JavidanEPJA}. The present work may also lead to the formulation of a nonadditive equation of state that may be employed in studies of nonlinear waves inside Quark-Gluon Plasma \cite{tbepjc1}, or studying stellar matter \cite{deppmanepja}. 

\section*{Acknowledgements}
TB acknowledges funding from the European Union's HORIZON EUROPE programme, via the ERA Fellowship Grant Agreement number 101130816. GW was supported in part by the Polish Ministry of Education and Science, Grant No. 2022/WK/01.

\end{document}